\documentclass[a4paper,12pt]{article}

\usepackage{soul}
\usepackage[usenames,dvipsnames]{color}
\usepackage{jheppub}
\usepackage{amsmath, amssymb, slashed, epsf, color, graphicx, latexsym}
\usepackage{epsfig}
\usepackage{color}

\begin{document}
	\preprint{}
	\title{de Sitter Static Black Ring in Large $D$ Membrane Paradigm at the Second Order}

	\author{Mangesh Mandlik}
	
	\affiliation{Department of Physics, Indian Institute of Technology, Kanpur 208016, India.}

	\emailAdd{mangeshm@iitk.ac.in}

	\abstract{It was shown in \cite{Mandlik:2020lgf} that the effective stationary membrane equations from the large $D$ membrane paradigm at the first order admit black ring solutions in flat and AdS cases, but the de Sitter solution obtained in \cite{Caldarelli:2008pz} lies outside the domain of their applicability. In this short note the static de Sitter black ring is obtained from the second order membrane paradigm, and it satisfies the equilibrium condition for the thin ring solution of \cite{Caldarelli:2008pz}. This provides a segue into the stationary black rings at the second order.}
	
	\maketitle
	
\section{Introduction}\label{intro}

Building on the observation by Emparan, Suzuki, Tanabe and collaborators \cite{Emparan:2013moa, Emparan:2013xia, Emparan:2014cia, Emparan:2014jca, Emparan:2014aba} that the black hole dynamics simplifies tremendously when the number of spacetime dimensions $D$ is taken to be large, a program called ``Large $D$ membrane paradigm" was first successfully developed in \cite{Bhattacharyya:2015dva}. In this program the dynamics of a $D$ dimensional black hole is mapped onto the dynamics of a $D-1$ dimensional hypersurface that does not backreact on its $D$ dimensional background spacetime in which it moves. This allows to solve for a particular class of black holes systematically in perturbative orders of $1/D$ which becomes small as $D$ gets large, making the approximation better. Although in \cite{Bhattacharyya:2015dva} it was developed for the asymptotically flat vacuum solutions only to the first subleading order in $1/D$, later it was extended to Einstein-Maxwell system, including cosmological constant and to second order (see \cite{Bhattacharyya:2015fdk, Dandekar:2016fvw, Bhattacharyya:2017hpj, Bhattacharyya:2018szu, Kundu:2018dvx}).

In \cite{Mandlik:2018wnw} the first order membrane paradigm was restricted to stationary configurations, and the effective equations that such configurations obey were obtained. These equations were solved for axisymmetric configurations, and they were found to admit asymptotically flat stationary solutions with horizon topology $S^1\times S^{D-3}$. This class of solutions was discovered in \cite{Emparan:2001wk, Emparan:2001wn} for $D=5$ and then for any $D>4$ \cite{Emparan:2007wm} within `thin ring' approximation. Later in \cite{Mandlik:2020lgf}  The asymptotically AdS black rings were found from the first order stationary membrane paradigm, and also the equilibrium rotational speed of the asymptotically flat black ring was determined from the second order equation, which matched with that reported in \cite{Caldarelli:2008pz}.

However, \cite{Caldarelli:2008pz} also has an asymptotically de Sitter black ring solution which is static at an equilibrium `ring radius'. This solution was absent from the set of solutions of the stationary membrane paradigm of \cite{Mandlik:2020lgf} because the parameters for the reported solution lied outside the domain of validity of \cite{Mandlik:2018wnw}. The author indicated that consideration of the second order membrane paradigm should recover this solution, and this note does exactly that.

The membrane paradigm with cosmological constant \cite{Bhattacharyya:2017hpj, Bhattacharyya:2018szu} aims at finding black hole solutions for
\begin{equation}
R_{AB} - \frac{R}2G_{AB} +\frac{(D-1)(D-2)}2\lambda G_{AB} = 0.
\end{equation}
The `scaled' cosmological constant $\lambda$ is positive for asymptotically de Sitter solutions and has dimensions of inverse square of length. So we can write $\lambda = 1/L^2$ where $L$ is the characteristic dS length scale.

The setup of this membrane paradigm sets the length scale of the horizon of the solution to $\mathcal{O}\left(D^0\right)$. Then it assumes that $L \sim \mathcal{O}\left(D^0\right)$. That's why the first order can't detect the solutions where there is a parametric difference between the length scales of the black hole and the background de Sitter. The static dS solution of \cite{Caldarelli:2008pz} has $b/L \sim \mathcal{O}\left(D^{-1/2}\right)$, where $b$ is the $S^1$ radius of the ring or the `ring radius'. So we need to rescale the dS length scale as 
$$L^2 = D~\tilde{L}^2,$$
where $\tilde{L}\sim\mathcal{O}(D^0)$. But this length scale will be significant only if we go to the second order. The rest of this paper explores this idea.

\section{The calculation}\label{calc}
To recap \cite{Mandlik:2020lgf}, for stationary axisymmetric configurations the $(t,r,\theta,s,\{\chi^a\})$ coordinate system, fondly known as $r-s$ coordinate system, is constructed by splitting the background $R^{D-1,1}$ into $R\times R^2 \times R^{D-3}$ where $R$ is the time direction parametrized by $t$ and $R^2$ is the plane of axisymmetry. $(r,\theta)$ are the polar coordinates on the $R^2$ while $(s,\{\chi^a\})$ are the spherical polar coordinates in the remaining $R^{D-3}$. The background de Sitter metric becomes
\begin{equation}\label{dSmetrs}
\begin{split}
G_{tt} &= -\left(1-\frac{r^2+s^2}{L^2}\right),\\
G_{rr} &= \frac{L^2-s^2}{L^2-(r^2+s^2)},\\
G_{rs} &= \frac{rs}{L^2-(r^2+s^2)},\\
G_{ss} &= \frac{L^2-r^2}{L^2-(r^2+s^2)},\\
G_{\theta\theta} &= r^2,\\
G_{ab} &= s^2\hat{g}_{ab},
\end{split}
\end{equation}
where $\hat{g}_{ab}$ is the metric on a unit $S^{D-4}$ in the coordinates $\{\chi^a\}$.\\
Considering the symmetries of this problem, the shape of the membrane corresponding to an axisymmetric stationary configuration would be given by the equation
$$s^2 = 2g(r).$$
The membrane (horizon) topology can be read off this equation. The `black hole' type solutions ($S^{D-2}$) have the maximum at $r = 0$. For example, the singly rotating black holes have the form $2g = R^2 - a r^2$. The `black ring' type solutions $S^1\times S^{D-3}$ necessarily have the maximum at $r>0$, e.g. $2g = R^2 - (r-b)^2$.\\

For a static configuration the membrane velocity is given by
$$u = \gamma k,$$
where $k \equiv \partial_t$ is the Killing vector of the background that corresponds to static configurations. Due to the normalization condition $G_{MN}u^Mu^N = -1$, we get
\begin{equation}
\gamma = \sqrt{-G_{tt}} = \left(1-\frac{r^2+s^2}{L^2}\right)^{-1/2}
\end{equation}
At the first subleading order in $D$, the membrane equation is \cite{Mandlik:2018wnw}
$$K = \frac{D\gamma}{\beta},$$
where $K$ is the trace of the extrinsic curvature of the membrane as embedded in the background de Sitter, and $\beta = \frac{D}{4\pi T}$, $T$ being the temperature.\\
In \cite{Mandlik:2020lgf} the membrane equation was ultimately cast in the form (for static uncharged case $\omega = 0$ and $\alpha = 0$, and analytically continuing $L \to iL$ for de Sitter),
\begin{equation}\label{stmemeq}
2g+\left(g'\right)^2-\frac{\left(2g-rg'\right)^2}{L^2}=\beta^2\left(1-\frac{2g-rg'}{L^2}\right)^2\left(1-\frac{2g+r^2}{L^2}\right),
\end{equation}
where prime denotes a derivative w.r.t. $r$.

As shown in \cite{Mandlik:2020lgf}, this equation dictates that $g$ can have a maximum only at $r=0$, which indicates a horizon topology $S^{D-2}$ , not $S^1 \times S^{D-3}$. Hence there can't be a static dS black ring solution at this order.

However, note that if $L^2$ is scaled up by $\mathcal{O}(D)$, at the leading order the metric \eqref{dSmetrs} becomes flat and \eqref{stmemeq} reduces to
$$2g+\left(g'\right)^2 = \beta^2,$$
which is the same as the equation for static uncharged membrane in flat background in $r-s$ coordinates, and according to \cite{Mandlik:2020lgf} does admit a black ring solution
$$2g = \beta^2 - (r-b)^2,$$
$b$ is called the `ring radius'.

Now the effect of the ``de Sitter-ness" of the metric can be treated as a $1/D$ perturbation. Then we need to take into account the second order correction to the membrane equations. The second order uncharged membrane equation is given in \cite{Bhattacharyya:2018szu}:
\begin{equation}\label{2ndorder}
\begin{split}
&\color[rgb]{0.00,0.00,1.00}{\left[\frac{\nabla^2 u_\alpha}{ K}+u^\beta { K}_{\beta\alpha}-\frac{u^\beta { K}_{\beta \delta} { K}^\delta_\alpha}{ K}-\nabla_\alpha\ln K-u\cdot\nabla u_\alpha\right]P^\alpha_\gamma}\\
&
+ \Bigg[\frac{\nabla^2\nabla^2 u_\alpha}{{ K}^3}-\frac{(\nabla_\alpha{ K})(u\cdot\nabla{ K})}{{ K}^3}-\frac{(\nabla_\beta{ K})(\nabla^\beta u_\alpha)}{{ K}^2}-\frac{2{ K}^{\delta \sigma}\nabla_\delta\nabla_\sigma u_\alpha}{K^2} -\frac{\nabla_\alpha\nabla^2{ K}}{{ K}^3}
\\&+\frac{\nabla_\alpha({ K}_{\beta\delta} { K}^{\beta\delta} { K})}{K^3}+3\frac{(u\cdot { K}\cdot u)(u\cdot\nabla u_\alpha)}{{ K}}\\
&-3\frac{(u\cdot { K}\cdot u)(u^\beta { K}_{\beta\alpha})}{{ K}}
-6\frac{(u\cdot\nabla{ K})(u\cdot\nabla u_\alpha)}{{ K}^2}+6\frac{(u\cdot\nabla{ K})(u^\beta { K}_{\beta\alpha})}{{ K}^2}+3\frac{u\cdot\nabla u_\alpha}{D-3}\\
&-3\frac{u^\beta { K}_{\beta\alpha}}{D-3}
-\frac{(D-1)}{K^2L^2}\left(\frac{\nabla_\alpha { K}}{ K}-2u^\sigma { K}_{\sigma\alpha}+2(u\cdot\nabla)u_\alpha\right)\Bigg]{\cal P}^\alpha_\gamma 
=0,
\end{split}
\end{equation}
where $K_{\alpha\beta}$ are the extrinsic curvature components of the membrane as embedded in the background, and $P^{\alpha}_{\gamma}$ is the projector orthogonal to the velocity,
$$P^{\alpha}_{\gamma} = \delta^\alpha_\gamma + u^\alpha u_\gamma.$$
The Greek indices denote the coordinates on the membrane. The covariant derivatives and the raising and lowering of indices are w.r.t. the induced metric $g_{\alpha\beta}$ on the membrane. The leading order piece of the equation is coloured blue.\\ 
In appendix \ref{simp} this equation is shown to simplify a lot for static configurations and becomes
$$K = \frac{\gamma}{\epsilon\beta},$$
where $\epsilon = \frac{1}{D-4}$ is the perturbation parameter. This time $K$ and $\gamma$ receive the corrections due to the $1/L^2$ terms in the metric. This equation is solved in appendix \ref{solapp}. Demanding the solution is regular imposes a condition
$$b = \frac{L}{\sqrt{D}},$$
which is precisely the condition that the static dS black ring of \cite{Caldarelli:2008pz} satisfies.

\section{Discussion}\label{disc}

This short note complements the result of \cite{Mandlik:2020lgf} by obtaining the static de Sitter black ring. Though the large $D$ membrane paradigm is constructed by assuming the black hole membrane and the embedding (A)dS spacetime have parametrically similar length scales, it is also capable of delivering results beyond that assumption, in the direction where (A)dS length scales are larger. This sounds familiar, as in \cite{Mandlik:2020lgf} it was shown that membrane velocities that are parametrically slower than order unity $D^0$ also be analysed within the scope of stationary membrane paradigm. It remains to check whether membrane paradigm suffers a breakdown for dynamical configurations outside its `design' parameter regime.

The immediate follow up to this work would be to obtain second order stationary membrane equations in the same way the first order membrane paradigm was restricted to the stationary configurations in \cite{Mandlik:2018wnw}. This will help find newer stationary solutions, including rotating de Sitter black rings, and also explore new features of the first order solutions. It would be interesting to do stability analysis of such solutions by calculating their quasinormal modes. Along with the singly rotating black holes, this will help in constructing the phase diagram of axisymmetric black holes at large $D$.

\section*{Acknowledgements}
I would like to acknowledge my debt to the people of India for their generous and steady support to research in the basic sciences.

\appendix
\section{Simplification of the membrane equations}\label{simp}
Let's recall the second order uncharged membrane equation of motion \eqref{2ndorder}
\begin{equation}
\begin{split}
&\mathcal{E}_\gamma\equiv\color[rgb]{0.00,0.00,1.00}{\left[\frac{\nabla^2 u_\alpha}{ K}+u^\beta { K}_{\beta\alpha}-\frac{u^\beta { K}_{\beta \delta} { K}^\delta_\alpha}{ K}-\nabla_\alpha\ln K-u\cdot\nabla u_\alpha\right]P^\alpha_\gamma}\\
&
+ \Bigg[\frac{\nabla^2\nabla^2 u_\alpha}{{ K}^3}-\frac{(\nabla_\alpha{ K})(u\cdot\nabla{ K})}{{ K}^3}-\frac{(\nabla_\beta{ K})(\nabla^\beta u_\alpha)}{{ K}^2}-\frac{2{ K}^{\delta \sigma}\nabla_\delta\nabla_\sigma u_\alpha}{K^2} -\frac{\nabla_\alpha\nabla^2{ K}}{{ K}^3}
\\&+\frac{\nabla_\alpha({ K}_{\beta\delta} { K}^{\beta\delta} { K})}{K^3}+3\frac{(u\cdot { K}\cdot u)(u\cdot\nabla u_\alpha)}{{ K}}\\
&-3\frac{(u\cdot { K}\cdot u)(u^\beta { K}_{\beta\alpha})}{{ K}}
-6\frac{(u\cdot\nabla{ K})(u\cdot\nabla u_\alpha)}{{ K}^2}+6\frac{(u\cdot\nabla{ K})(u^\beta { K}_{\beta\alpha})}{{ K}^2}+3\frac{u\cdot\nabla u_\alpha}{D-3}\\
&-3\frac{u^\beta { K}_{\beta\alpha}}{D-3}
-\frac{(D-1)}{K^2L^2}\left(\frac{\nabla_\alpha { K}}{ K}-2u^\sigma { K}_{\sigma\alpha}+2(u\cdot\nabla)u_\alpha\right)\Bigg]{\cal P}^\alpha_\gamma 
=0.
\end{split}
\end{equation}
From \eqref{dSmetrs} we have $G_{tM} = 0 ~\forall M\neq t$, and all $G_{MN}$ are time independent. Also, the shape of the membrane is time independent. So the membrane can be parametrized by the coordinates $(t, {y^p})$ so that $g_{pt} = 0$ and all $g_{\mu\nu}$ are independent of $t$. The projection of velocity on the membrane is trivial, $u = \gamma\partial_t$. So
$$P^t_\mu = 0 = P^\mu_t,$$
which means $\mathcal{E}_t$ = 0 identically. So we analyse $\mathcal{E}_p$.\\
Now, the independence of the metric and the shape function on $t$ and vanishing of $G_{tM}$ for $M\neq t$ gives $K_{tp} = 0$. So
\begin{equation}\label{van1}
u^\beta K_{\beta p} = 0,~~~ u^\beta K_{\beta \delta}K^\delta_p = 0. 
\end{equation}
Also, for a vector $V$ along $\partial_t$ (i.e. $V^p = 0~\forall p$),
$$\nabla^2 V_p = 0.$$
This comes from vanishing of $\Gamma^t_{pq}$ and $\Gamma^p_{tq}$. Its immediate consequence is $\nabla^2u_p = 0$, but this also implies $\nabla^2\nabla^2u_p = 0$ by putting $V = \nabla^2u$.\\
Also, as shown in \cite{Mandlik:2018wnw}, for a stationary velocity $u$,
$$u\cdot\nabla u_\mu = -\partial_\mu\ln\gamma.$$
Thus in the static regime the second order membrane equation simplifies to
\begin{equation}\label{simpstat}
\begin{split}
&\mathcal{E}_p\equiv\color[rgb]{0.00,0.00,1.00}{-\partial_p\ln\left(\frac{K}{\gamma}\right)}\\
&-\frac{(\nabla_p{ K})(u\cdot\nabla{ K})}{{ K}^3}-\frac{(\nabla_\beta{ K})(\nabla^\beta u_p)}{{ K}^2}-\frac{2{ K}^{\delta \sigma}\nabla_\delta\nabla_\sigma u_p}{K^2} -\frac{\nabla_p\nabla^2{ K}}{{ K}^3}
\\&+\frac{\nabla_p({ K}_{\beta\delta} { K}^{\beta\delta} { K})}{K^3}-3(\partial_p\ln\gamma)\left(\frac{u\cdot { K}\cdot u}{{ K}}-2\frac{(u\cdot\nabla{ K})}{{ K}^2}+\frac{1}{D-3}\right)\\
&-\frac{D-1}{K^2L^2}\left(\partial_p\ln K-2\partial_p\ln\gamma\right)=0.
\end{split}
\end{equation}
Now look at the terms in black. As we will see in the next appendix \ref{solapp}, $K$ and $\gamma$ are constant at the leading order in $1/D$ (or $\epsilon$). Thus all the terms in black involving derivatives of $K$ and $\gamma$ become subleading and drop out. Also, with the new scaling of $L^2$ the terms involving $1/L^2$ explicitly also become subleading and drop out.\\
It can be shown that at leading order
\begin{equation}\label{van2}
\begin{split}
K^{\gamma\delta}\nabla_\gamma\nabla_\delta u_p &= \frac{K}{D}\nabla^2u_p,\\
K^{\gamma\delta}K_{\gamma\delta} &= \frac{K^2}{D},
\end{split}
\end{equation}
Which makes remaining terms in black subleading too. So finally the membrane equation reduces to
\begin{equation}
\partial_p\ln\left(\frac{K}{\gamma}\right) = 0,
\end{equation}
which can be integrated to get
\begin{equation}\label{netsimp}
K = \frac{\gamma}{\epsilon\beta},
\end{equation}
where $\epsilon \equiv \frac{1}{D-4}$.

\section{Detailed solution}\label{solapp}
Let's define the rescaled de Sitter length scale as
$$L^2 = \frac{\tilde{L}^2}{\epsilon},$$
so the background metric \eqref{dSmetrs} becomes
\begin{equation}\label{dSmetresc}
\begin{split}
G_{tt} &= -\left(1-\epsilon\frac{r^2+s^2}{\tilde{L}^2}\right),\\
G_{rr} &= 1+\epsilon\frac{r^2}{\tilde{L}^2},\\
G_{rs} &= \epsilon\frac{rs}{\tilde{L}^2},\\
G_{ss} &= 1+\epsilon\frac{s^2}{\tilde{L}^2},\\
G_{\theta\theta} &= r^2,\\
G_{ab} &= s^2\hat{g}_{ab},
\end{split}
\end{equation}
with the inverse metric
\begin{equation}\label{dSmetinvresc}
\begin{split}
G^{tt} &= -\left(1+\epsilon\frac{r^2+s^2}{\tilde{L}^2}\right),\\
G^{rr} &= 1-\epsilon\frac{r^2}{\tilde{L}^2},\\
G^{rs} &= -\epsilon\frac{rs}{\tilde{L}^2},\\
G^{ss} &= 1-\epsilon\frac{s^2}{\tilde{L}^2},\\
G^{\theta\theta} &= \frac{1}{r^2},\\
G^{ab} &= \frac{1}{s^2}\hat{g}^{ab}.
\end{split}
\end{equation}
The determinant of the metric is
$$G = -r^2s^{2(D-4)}\Omega^2_{D-4},$$
where $\Omega_{D-4}$ is the volume measure of $S^{D-4}$.
We begin with the simplified equation \eqref{netsimp}
\begin{equation}
K = \frac{\gamma}{\epsilon\beta},
\end{equation}
and we keep only the terms of $\mathcal{O}\left(\epsilon^{-1}\right)$ and $\mathcal{O}\left(\epsilon^{0}\right)$ on both sides.\\
Also we define the perturbed shape function
\begin{equation}
2g = 2g_0 + 2\epsilon\mathfrak{g},
\end{equation}
where the unperturbed static black ring shape function is given by 
\begin{equation}
2g_0 = \beta^2 - (r-b)^2,
\end{equation}
since it solves the leading order equation.

The normal to the membrane is $n = n_r dr + n_s ds$ where
\begin{equation}
n_s = \mathcal{N}s,~~~~n_r = -\mathcal{N}g' = \mathcal{N}\left((r-b) - \mathfrak{g}'\right).
\end{equation}
$\mathcal{N}$ is the normalization chosen so that $G^{AB}n_An_B = 1$. So
\begin{equation}
\begin{split}
\mathcal{N}^{-2} &= 2g + (g')^2 - \frac{\epsilon}{\tilde{L}^2}\left(2g-rg'\right)^2\\
&= \beta^2+2\epsilon\left(\mathfrak{g}-(r-b)\mathfrak{g}'\right)-\frac{\epsilon}{\tilde{L}^2}\left(\beta^2+b(r-b)\right)^2.
\end{split}
\end{equation}
which gives
\begin{equation}
\mathcal{N} = \frac{1}{\beta}\left(1-\frac{\epsilon}{\beta^2}\left(\mathfrak{g}-(r-b)\mathfrak{g}'\right)+\frac{\epsilon}{2\beta^2\tilde{L}^2}\left(\beta^2+b(r-b)\right)^2\right)
\end{equation}

Due to the explicit $1/\epsilon$ factor on the RHS we keep $\gamma$ up to $\mathcal{O}\left(\epsilon^{1}\right)$
\begin{equation}\label{gamexp}
\gamma = 1 + \frac{\epsilon}2\frac{2g+r^2}{\tilde{L}^2} = 1 + \frac{\epsilon}2\frac{\beta^2+r^2-(r-b)^2}{\tilde{L}^2},
\end{equation}
while on LHS,
\begin{equation}\label{Kexp}
\begin{split}
K &= \frac{1}{\sqrt{-G}}\partial_M\left(\sqrt{-G}n^M\right),\\
&= \frac{n^s}{\epsilon s}+\frac{n^r}{r}+\partial_sn^s+\partial_rn^r.
\end{split}
\end{equation}
We are interested in terms of orders $\epsilon^{-1}$ and $\epsilon^{0}$. So only the term $\frac{n^s}s$ needs to be calculated up to $\mathcal{O}\left(\epsilon^1\right)$
\begin{equation}\label{nr}
n^r = G^{rr}n_r+G^{rs}n_s = -\mathcal{N}g'+\mathcal{O}\left(\epsilon^1\right) = \frac{r-b}{\beta}+\mathcal{O}\left(\epsilon^1\right),
\end{equation}

\begin{equation}\label{ns}
\begin{split}
n^s &= G^{rs}n_r+G^{ss}n_s = \mathcal{N}s - \epsilon\frac{\mathcal{N}s}{\tilde{L}^2}\left(2g-rg'\right)+\mathcal{O}\left(\epsilon^2\right),\\
&= \frac{s}{\beta}\left(1-\frac{\epsilon}{\beta^2}\left(\mathfrak{g}-(r-b)\mathfrak{g}'\right)-\epsilon\frac{\beta^2}{2\tilde{L}^2}+\epsilon\frac{b^2(r-b)^2}{2\beta^2\tilde{L}^2}\right),
\end{split}
\end{equation}

\begin{equation}\label{der}
\partial_rn^r+\partial_sn^s = \frac{2}{\beta}+\mathcal{O}\left(\epsilon^1\right),
\end{equation}
Thus from \eqref{Kexp}, \eqref{ns}, \eqref{nr} and \eqref{der},
\begin{equation}\label{Kfin}
K = \frac{1}{\beta}\left(\frac{1}\epsilon-\frac{\mathfrak{g}-(r-b)\mathfrak{g}'}{\beta^2}-\frac{\beta^2}{2\tilde{L}^2}+\frac{b^2(r-b)^2}{2\beta^2\tilde{L}^2}+2+\frac{r-b}{r}\right).
\end{equation}
Hence \eqref{netsimp} becomes using \eqref{gamexp} and \eqref{Kfin},
\begin{equation}
(r-b)\mathfrak{g}'-\mathfrak{g} = \frac{\beta^4}{\tilde{L}^2}-2\beta^2-\beta^2\left(\frac{r-b}{r}\right)+\frac{\beta^2r^2}{2\tilde{L}^2} - \frac{(b^2+\beta^2)(r-b)^2}{2\tilde{L}^2}.
\end{equation}
This is a first order ODE which can be easily solved to get
\begin{equation}
\begin{split}
\mathfrak{g}(r) =& 2\beta^2- \frac{\beta^4}{\tilde{L}^2} + \frac{\beta^2(r-b)}{b}\ln r-\frac{\beta^2b^2}{2\tilde{L}^2}-\frac{b^2}{2\tilde{L}^2}r(r-b)\\
&+\beta^2\left(\frac{b}{\tilde{L}^2}-\frac{1}b\right)(r-b)\ln(r-b) +C(r-b),
\end{split}
\end{equation}
Where $C$ is the integration constant. The solution is nonsense for $r<b$ unless the coefficient of $\ln(r-b)$ vanishes. This gives the condition
\begin{equation}
b = \tilde{L}.
\end{equation}
\newpage
\bibliography{dSBR}

\end{document}